\documentclass[12pt]{article}
\usepackage{amsfonts,amsmath}
\usepackage{amssymb}
\usepackage[usenames]{color}
\usepackage{mathrsfs}
\usepackage{cancel}
\usepackage{tikz}
\usepackage{soul}
\usepackage{bbm}
\usepackage[T2A]{fontenc}
\usepackage{graphicx}
\usepackage[utf8]{inputenc}

\newcommand{\dr}{{{\rm d}}}
\renewcommand{\theequation}{\thesection.\arabic{equation}}

\makeatletter \@addtoreset{equation}{section} \makeatother

{\vspace{3mm} }

\def\*{\star}

\def\E2{\mathbf{E}}

\newcommand{\be}{\begin{equation}}
\newcommand{\ee}{\end{equation}}
\newcommand{\bee}{\begin{eqnarray}}
\newcommand{\beee}{\begin{array}}
\newcommand{\eee}{\end{eqnarray}}
\newcommand{\eeee}{\end{array}}
\newcommand{\gga}{\gamma}

\newcommand{\go}{\omega}

\newcommand{\half}{\frac{1}{2}}

\newcommand{\hmt}{\vartriangle}

\begin{document}

\begin{flushright}
\end{flushright}

\vspace{0.5cm}
\begin{center}
{\large\bf Gauge transformations in Z-space in (anti)holomorphic sector of HS theory}

\vspace{1 cm}

\textbf{D.A.~Batyaev, A.V.~Korybut}\\

\vspace{1 cm}

\textbf{}\textbf{}\\
 \vspace{0.5cm}
 \textit{I.E. Tamm Department of Theoretical Physics,
Lebedev Physical Institute,}\\
 \textit{ Leninsky prospect 53, 119991, Moscow, Russia }\\
\vspace{0.7 cm} batiaev.da@phystech.edu, akoribut@gmail.com
\par\end{center}

\begin{center}
\vspace{0.6cm}

\par\end{center}

\vspace{0.4cm}

\begin{abstract}
\noindent 

\noindent We consider a consistent deformation of the (anti)holomorphic generating system of \cite{Didenko:2022qga}, aiming to provide a map to the (anti)holomorphic truncation of the Vasiliev theory. In our new formulation, the previously rigidly defined  master-field $\Lambda$ can be shifted by $\dr_z$-exact projective one-forms. The class of projective one-forms is described comprehensively, and corresponding projective identities are proven. $\dr_z$-exact forms of the order $n$ in $C$ induce field redefinition of order $(n+1)$ and beyond. Explicit expression for the $(n+1)$-th order field redefinitions are provided. An analysis of the gauge transformation of the second order in $C$ in zero-forms is performed, identifying necessary and sufficient conditions for the part of the field redefinition of the second order that can be gauged away. The field redefinitions induced by the shift of $\Lambda$ are proven to be nontrivial.
\end{abstract}
\newpage

\newpage

\tableofcontents
\section{Introduction}
Higher spin gauge theory is a field theory describing gauge interactions between gauge fields of all spins. Theories possessing such a large symmetry algebra are argued to describe interactions at ultra-high energies \cite{Vasiliev:2016xui} when all symmetries become manifest. Also, due to the large symmetry algebra and hence the infinite spectrum, such theories find applications in holography \cite{Klebanov:2002ja} and cosmology \cite{Barvinsky:2015wvz}. Despite numerous yet unsolved problems like HS symmetry breaking (see, for example, \cite{Didenko:2023txr}, \cite{Vasiliev:2018zer}, \cite{Silva:2021ece}, \cite{Alday:2015ota}, \cite{Maldacena:2012sf} for recent attempts), there is still an unresolved question regarding locality. There are early observations that show that HS theory cannot be a local field theory in the traditional sense; the number of derivatives in interaction vertices grows with spin indefinitely \cite{Bengtsson:1983pd}, \cite{Berends:1984wp}, \cite{Fradkin:1987ks}, \cite{Fradkin:1991iy}. On the other hand, it is precisely the HS algebra and its spectrum that allow us to cancel infinities in pioneering quantum models \cite{Giombi:2013fka}, \cite{Beccaria:2014xda}.

While the traditional notion of locality is not applicable, a different one was developed, the so-called {\it spin-locality}. Such locality implies that for a given set of spins, the interaction vertex contains only a finite number of derivatives. It is worth highlighting that the particular local functional class to which the vertices belong is vital for the consistent formulation of the theory. Constraining the functional class for the vertices, one also constrains possible field redefinitions and thus the remaining predictability of the theory.

There are many approaches to HS dynamics available in the literature (see recent reviews \cite{Bekaert:2004qos}, \cite{Didenko:2014dwa}, \cite{Ponomarev:2022vjb} and references therein). We confine ourselves to the so-called unfolded approach \cite{Vasiliev:1988xc}, \cite{Vasiliev:1988sa} (see \cite{Misuna:2026bhy},\cite{Zinoviev:2026osh},\cite{Iazeolla:2025btr} for recent applications of the formalism) and to the formalism of generating systems originally proposed by Vasiliev in \cite{Vasiliev:1990en} (see also \cite{Vasiliev:1992av}). The Vasiliev generating system allows us to write down the HS dynamics in the form
\begin{multline}\label{sch1form}
\dr_x \go+\go\ast \go=\Upsilon^\eta(\go,\go,C)+\Upsilon^{\bar{\eta}}(\go,\go,C)+\\
+\Upsilon^{\eta\eta}(\go,\go,C,C)+\Upsilon^{\eta\bar{\eta}}(\go,\go,C,C)+\Upsilon^{\bar{\eta}\bar{\eta}}(\go,\go,C,C)+\ldots ,
\end{multline}
\begin{multline}\label{sch0form}
\dr_x C+[\go,C]\ast=\Upsilon^\eta(\go,C,C)+\Upsilon^{\bar{\eta}}(\go,C,C)+\\+\Upsilon^{\eta\eta}(\go,C,C,C)+\Upsilon^{\eta\bar{\eta}}(\go,C,C,C)+\Upsilon^{\bar{\eta}\bar{\eta}}(\go,C,C,C)+\ldots
\end{multline}
Fronsdal double traceless spin $s$ field  $\varphi_{\nu_1\ldots \nu_s}(x)$
and its on-shell derivatives are packed into fields $\go$ and $C$
\begin{equation}
\go(y,\bar{y}|x)=\sum_{m,n}\dr x^\mu \omega_{\mu, \alpha_1 \ldots \alpha_m,\dot{\alpha}_1 \ldots \dot{\alpha}_n} y^{\alpha_1}\ldots y^{\alpha_m}\bar{y}^{\dot{\alpha}_1}\ldots \bar{y}^{\dot{\alpha}_n}\;\;\; m+n=2(s-1)\,,
\end{equation}
\begin{equation}
C(y,\bar{y}|x)=\sum_{m,n}C_{\alpha_1 \ldots \alpha_m,\dot{\alpha}_1\ldots \dot{\alpha}_n} y^{\alpha_1}\ldots y^{\alpha_m}\bar{y}^{\dot{\alpha}_1}\ldots \bar{y}^{\dot{\alpha}_n}\;\;\; |m-n|=2s\,.
\end{equation}
Both functions $\go$ and $C$ are endowed with the associative Moyal-product
\begin{equation}\label{Moyal}
f(y,\bar{y})\ast g(y,\bar{y})=f(y,\bar{y})\exp\left\{i\epsilon^{\alpha \beta} \frac{\overleftarrow{\partial}}{\partial y^\alpha}\frac{\overrightarrow{\partial}}{\partial y^\beta}+i\bar{\epsilon}^{\dot{\alpha}\dot{\beta}}\frac{\overleftarrow{\partial}}{\partial \bar{y}^{\dot{\alpha}}}\frac{\overrightarrow{\partial}}{\partial \bar{y}^{\dot{\beta}}}\right\}g(y,\bar{y})\,.
\end{equation}
Linear perturbations over the $AdS_4$ background correspond to the free Fronsdal equations \cite{Fronsdal:1978rb}, this correpondence is the content of the famous central on-mass-shell theorem \cite{Vasiliev:1999ba} (see also \cite{Bychkov:2021zvd}). $\eta$ and $\bar{\eta}$ are the coupling constants of the theory stored in the interaction vertices $\Upsilon$.

Equations \eqref{sch1form}, \eqref{sch0form} can be equivalently rewritten in the metric-like formalism
\begin{multline}
\Box \varphi^{\mu_1\ldots \mu_s}-\nabla^{\mu_1}\nabla_\nu \varphi^{\nu\mu_2\ldots \mu_s}+\frac{1}{2}\nabla^{\mu_1}\nabla^{\mu_2}\varphi^{\mu_3\ldots \mu_s \nu} {}_\nu-m_s^2\varphi^{\mu_1\ldots \mu_s}+2\Lambda g^{\mu_1\mu_2}\varphi^{\mu_3\ldots \mu_s\nu} {}_\nu=\\
=\mathcal{V}^{\mu_1\ldots\mu_s}(\varphi,\varphi)+\mathcal{V}^{\mu_1\ldots\mu_s}(\varphi,\varphi,\varphi)+\ldots
\end{multline}
providing a nonlinear correction to free Fronsdal dynamics,  where mass-like parameter depends on spin $m_s^2=\Lambda\big(s-(s-2)(s+1)\big)$. Cubic (at the action level) interaction vertices in this metric-like form were obtained from the unfolded vertices in \cite{Misuna:2017bjb}, \cite{Tatarenko:2024csa}. It is worth mentioning that while cubic (at the action level) corrections were completely classified \cite{Metsaev:2005ar} (see also \cite{Metsaev:2007rn}), higher-order interaction vertices are mainly available in different formalisms. For example, from the Vasiliev framework, contributions to interaction vertices up to fifth order at the action level were obtained \cite{Didenko:2019xzz}, \cite{Didenko:2020bxd}, \cite{Gelfond:2021two}. For the vertices in the unfolded formalism to be turned into spin-local ones in the metric-like formalism, they are required to be {\it projectively compact} \cite{Vasiliev:2022med}. This property is fulfilled for all (anti)holomorphic vertices discovered within the Vasiliev framework. Introducing a slight abuse of notation, we would call the projectively compact vertices given in the unfolded form spin-local, or just local.

Up to field redefinitions, $\eta$ and $\bar{\eta}$ are the only coupling constants of the theory \cite{Vasiliev:1990en}. Vertices 
$\Upsilon^{\bullet\bullet}(\bullet,\ldots,\bullet)$ on the r.h.s. of \eqref{sch1form}, \eqref{sch0form} are proportional to the powers of 
$\eta$ and $\bar{\eta}$ indicated in the superscript. Vertices proportional exclusively to the powers of $\eta$ are called holomorphic; similarly, vertices proportional to powers of $\bar{\eta}$ are called antiholomorphic. Technically, in such vertices, the contraction between dotted indices is described exclusively by the dotted part of the product \eqref{Moyal} in the holomorphic sector, and vice versa in the antiholomorphic one. For that reason we omit writing barred variables explicitly and barred Moyal-product is implicit. The counterpart to pure holomorphic and antiholomorphic vertices is the so-called mixed sector, with vertices proportional to both $\eta$ and $\bar{\eta}$. This sector is far less investigated in the literature compared to the (anti)holomorphic one; nonetheless, some results are available. Introducing a milder notion of locality, the vertices $\Upsilon^{\eta\bar{\eta}}(\go,\go,C,C)$ and $\Upsilon^{\eta\bar{\eta}}(\go,C,C,C)$ were computed \cite{Didenko:2019xzz}, \cite{Gelfond:2023fwe} (see also \cite{Didenko:2026nag}, where a different generating system was developed to attack the mixed sector of HS theory). As shown in \cite{Vasiliev:1992av}, HS theory admits a consistent truncation leaving only, for example, holomorphic vertices non-trivial. Such theories are called {\it self-dual} HS theories or {\it chiral} in the literature.

The term {\it generating system} means that to obtain the equations \eqref{sch1form}, \eqref{sch0form}, one needs to solve additional equations. These equations describe the extra dependence of the so-called master-fields on the auxiliary variables $z$. The equations are designed in such a way that the dependence on these auxiliary $z$-variables vanishes in the resulting dynamics, and only the dependence on the $y$s survives in \eqref{sch1form}, \eqref{sch0form}.

Freedom of (anti)holomorphic truncation was successfully utilized in \cite{Didenko:2022qga} (see also \cite{Didenko:2023vna}, where a similar generating system was designed to describe symmetric HS fields in 
$d>4$), where a generating system that produces only (anti)holomorphic vertices was proposed. The generated vertices are proven to be local in all orders and were found explicitly \cite{Didenko:2024zpd} (see also \cite{Sharapov:2022nps} which is based upon \cite{Ponomarev:2016lrm}). Recently, the (anti)holomorphic system was generalized to also encompass the mixed sector \cite{Didenko:2026nag}, eventually covering the full HS dynamics. However, it is worth mentioning that the computation of mixed-sector interaction vertices from \cite{Didenko:2026nag} is not as straightforward as that in \cite{Vasiliev:1992av}; namely, on the way  to \eqref{sch1form},  \eqref{sch0form} one faces an additional equation which is absent in Vasiliev formulation. Nonetheless, it is worth mentioning that for the computation of $\Upsilon^{\eta\bar{\eta}}(\go,C,C,C)$, this equation was proven to be satisfied identically.

While there are several generating systems for HS dynamics, each with their own advantages and yet unresolved problems, the connection between them is not clear. The only nontrivial exact agreement available at the moment is for the vertices $\Upsilon^{\eta}(\go,C,C)$ \cite{Didenko:2022qga}, \cite{Korybut:2025vdn}; nonetheless, there are some indirect indications that the systems are indeed related, based on the shift‑symmetry properties of the vertices \cite{Didenko:2022eso}. It is argued in the literature that the self-dual system of \cite{Didenko:2022qga} can be obtained from the Vasiliev equations \cite{Vasiliev:1990en} by virtue of the $\beta\rightarrow-\infty$ limit, but such statements are far from being rigorous. Even though, in general, one may obtain all (anti)holomorphic interaction vertices \eqref{sch1form}, \eqref{sch0form} in both systems and look for field redefinitions of $\go$ and 
$C$ that relate them
\begin{equation}\label{field_redef}
\go\rightarrow\go+f_\go(\go,C,\ldots,C),\:\:\:\: C\rightarrow C+f_C(C,\ldots,C)
\end{equation}
such an approach does not look very promising from a technical point of view. In the first place, if two systems are related, then such a relation should be seen at the level of generating systems, which suggests looking at some deformation of either generating system. In this paper, we deform the (anti)holomorphic system proposed in \cite{Didenko:2022qga}. Its original formulation insisted on a rigid definition for some of its master-fields, namely $\Lambda[C]$, which should be a fixed functional of $C$. We relax this requirement, thus introducing a consistent deformation at the level of the generating system. More specifically, we allow the master-field 
$\Lambda$ to be an arbitrary solution to the corresponding inhomogeneous linear equation within a class of projective one-forms (to be defined later). This allows one to add various exact forms to a particular solution
\begin{equation}\label{gauge_Lambda}
\Lambda^\prime=\Lambda[C]+\dr_z\varepsilon,
\end{equation}
where  $\Lambda[C]$ is a rigidly defined functional from the original paper and $\dr_z \varepsilon$ should belong to the space of projective one-forms. We refer to \eqref{gauge_Lambda} as a {\it gauge transformation in 
$z$-space}, utilizing the traditional notion from electrodynamics\footnote{Similar shifts of the 
$S$ field are available for the Vasiliev system, but those shifts are allowed to be arbitrary $\dr_z$-exact forms.}. We would like to clarify an ambiguity that one may face when researching the literature. There is also a different notion of gauge transformation for the $\Lambda$-field available in the literature \cite{Faliakhov:2026yum} (see also \cite{Didenko:2023vna}). Even though the corresponding gauge parameters, say $\xi$, are allowed to be $z$-dependent, their effect on $\Lambda$ is constrained to be of the form
\begin{equation}\label{gauge_old}
\delta_\xi\Lambda[C]=\Lambda[\delta_\xi C]\,.
\end{equation}
It is obvious that such a gauge transformation has no effect on 
$\Lambda$ as a functional of $C$ and affects only 
$C$-field. Such a transformation does not change the interaction vertices resulting from the system.  In other words, one can consider gauge transformation of the unfolded generating system of \cite{Didenko:2022qga}, where $\Lambda$ is rigidly defined. For the $\Lambda$ field this gauge transformation is of the form \eqref{gauge_old}.  In contrast, the transformation of $\Lambda$ given by \eqref{gauge_Lambda} does change the generating system and hence does change the vertices. As anticipated in \cite{Didenko:2026nag}, it induces a field redefinition. We found the explicit expression for the second-order field redefinition induced by our deformation with $\varepsilon$ linear in $C$. Moreover, we considered 
$\varepsilon[C,\ldots,C]$ of the 
$n$th order in 
$C$, which induced field redefinitions of order 
$n+1$ and beyond, providing an explicit expression for the 
$(n+1)$-th order field redefinition. It is worth mentioning that each field redefinition \eqref{field_redef} is always given modulo gauge transformations, i.e., gauge transformations of eqs. \eqref{sch1form}, \eqref{sch0form}. For the second order in $C$, we found necessary and sufficient conditions for the part of the field redefinition that has no effect on the vertices and can be gauged away, providing explicitly the corresponding gauge parameter. It turns out that \eqref{gauge_Lambda} indeed produces a nontrivial deformation. Deformations of the kind \eqref{gauge_Lambda} were originally conjectured in \cite{Korybut:2025vdn}; however, the full description of projective one-forms was lacking at that time. Here we provide a description of this class\footnote{Class itself is the same as in \cite{Didenko:2026nag}.} in terms of a simple generating expression, which allows for writing the corresponding projective identities in a concise form, handy for practical computations. Moreover, $\dr_z$-exactness of the shift \eqref{gauge_Lambda} can be equivalently put in the form of a single integral constraint on the corresponding measure $\mu(\rho_1,\rho_2)$. This measure captures all the degrees of freedom of the field redefinition induced by the transformation \eqref{gauge_Lambda}.

Practical interest in the search for the relation between the Vasiliev system and the system of \cite{Didenko:2022qga} is the following. Even though one manages to get spin-local vertices from the Vasiliev generating equations, each new step required more and more sophisticated techniques\footnote{Development of various homotopy techniques for solving Vasiliev generating equations \cite{Gelfond:2018vmi}, \cite{Didenko:2018fgx}, \cite{Didenko:2019xzz} eventually led to the most general homotopy technique called {\it differential contracting homotopy} \cite{Vasiliev:2023yzx} (see \cite{Kirakosiants:2025gpd} for recent application). Nonetheless, it is not yet clear how to get, for example, all the (anti)holomorphic vertices in the spin-local form. It is worth mentioning that unconstrained field redefinitions are inherent in the Vasiliev system as the freedom of homogeneous solutions to the corresponding generating equations.}, in contrast, pure Poincaré lemma works perfectly for the system of \cite{Didenko:2022qga} in each order of the perturbation theory. Similarly, an all-order analysis exclusively by virtue of the Poincaré lemma was performed for the Vasiliev system \cite{Didenko:2015cwv}. Vertices obtained in this manner are non-local, which, however, does not imply that something is wrong with the generating equations as it was incorrectly interpreted in \cite{Skvortsov:2015lja}. Thus, having two systems solved by virtue of the Poincaré lemma, it is quite natural to look at the deformation of one that relates them, to utilize, for example, the success of the self-dual formulation. As a preliminary step of this quest, one should look at the current interaction, namely, at the vertices $\Upsilon^{\eta}(\go,C,C)$. In \cite{Vasiliev:2016xui}, by virtue of a field redefinition, this particular vertex from the pure Poincaré lemma approach in the Vasiliev framework was turned into a spin-local one. This vertex was extensively studied in the literature; for example, in \cite{Didenko:2017lsn} it was shown that it produces the correct boundary correlation function without introducing any sort of regularization, compared to holographic reconstruction from pure Poincaré  non-local version performed in \cite{Giombi:2009wh} (see also \cite{Giombi:2010vg}). Moreover, for the constrained spectrum, this is the only vertex that survives in self-dual gravity \cite{Didenko:2026apw} (see also \cite{Vasiliev:1989xz}). So far, we are not in a position to state if the change of the field frame from \cite{Vasiliev:2016xui} can be obtained as the field redefinition induced by our deformation. The main difficulty occurs due to the nonlinearity in $\varepsilon$ \eqref{gauge_Lambda} of the induced field redefinition.

The paper is organized as follows. In Section \ref{Burn_original} we recall the formulation of the (anti)holomorphic generating system of \cite{Didenko:2022qga}. We also show how interaction vertices are extracted up to the second order in $C$ in zero-forms. In Section \ref{projective_forms} we provide a description of the projective one-forms and identify which of them are the exact ones, which is equivalent to a simple integral constraint on the integration measure $\mu(\rho_1,\rho_2)$. In Section \ref{Burn_mango} we utilize the knowledge about projective one-forms, relaxing the rigid definition for $\Lambda$. We consider the gauge parameter $\varepsilon$ (cf. \eqref{gauge_Lambda}) linear in $C$ and extract the corresponding vertices up to the second order in $C$ in zero-forms. Section \ref{Field_redefinitions} is devoted to field redefinitions. We consider the generic case and provide a simple expression in terms of $\Lambda[C]$ and $\varepsilon[C]$ that relates the vertices obtained in Section \ref{Burn_original} and Section \ref{Burn_mango}. In Section \ref{Explicit_f}, using the description of the $\dr_z$-exact projective one-forms, we provide an explicit expression for the field redefinition in zero-forms induced by our deformation in terms of $\mu(\rho_1,\rho_2)$. We also show that such a deformation cannot be gauged away. Moreover, we show that the field redefinition induced by our deformation admits a local form for specific choices of the integration measures. The Conclusion contains a discussion of the obtained results as well as some open questions. In Appendix A, we present the proof of the projective identities for the one-forms described in Section \ref{projective_forms}. In Appendix B, we prove that the conditions promoted in Section \ref{Explicit_f} for the gauge part of the field redefinition are indeed necessary and sufficient.

\section{Generating system for (anti)holomorphic sector of HS theory}\label{Burn_original}

Generating system describing interaction in (anti)holomorphic sector proposed in \cite{Didenko:2022qga} looks as follows
\begin{align}
&\dr_{x} W(z,y|x)+W(z,y|x)*W(z,y|x)=0\,,\label{dxWeq1}\\
&\dr_z W(z,y|x)+\{W(z,y|x),\Lambda(z,y|x)\}_{*}+\dr_x\Lambda(z,y|x)=0\,,\label{Weq11}\\
&\dr_z\Lambda(z,y|x)=C(y|x)*\gga\,,\;\; \gamma:=\half\theta_\alpha\theta^\alpha e^{iz_\alpha y^\alpha}\,,\label{Lambdaeq1}\\
&\dr_x C(y|x)*\gga=\dr_z\{W(z,y|x),\Lambda(z,y|x)\}_{*}\,,\label{Ceq1}\\
&W(z,y|x)\in \mathbf{C}^0\, , \label{class1}\\
&\Lambda[C]:= \int_0^1 d\mathcal{T}\, \mathcal{T} \theta^\alpha z_\alpha e^{i\mathcal{T}z_\alpha y^\alpha}C(-\mathcal{T}z|x)\,. \label{LambdaDef}
\end{align}
Star-product used here 
\begin{equation}\label{star_infty}
(f\ast g)(z,y)
= \int \dr^2 u\,\dr^2 v \, \dr^2 P\, \dr^2Q\, e^{iu_\alpha v^\alpha -iP_\alpha v^\alpha+iu_\alpha Q^\alpha}f(z+u,y+P)g(z+v,y+Q)
\end{equation} 
generalizes Moyal product \eqref{Moyal} and will be referred to as {\it limiting star-product} since it emerges as a $\beta$-deformed product introduced in \cite{Didenko:2019xzz} (see also \cite{Sharapov:2022awp} where the same star-product was utilized). Differentials $\dr_x$ and $\dr_z$ are usual de Rham differential in spacetime and auxiliary $\theta^\alpha$ direction respectively
\begin{equation}
\dr_x:=\dr x^\mu \frac{\partial}{\partial x^\mu}\,,\;\; \dr_z:=\theta^\alpha \frac{\partial}{\partial z^\alpha}\,,\;\; \dr_x^2=\dr_z^2=\{\dr_x,\dr_z\}=0\,.
\end{equation}
One obtains dynamics \eqref{sch1form}, \eqref{sch0form} perturbatively (in powers of $C$) solving generating equations that describe evolution in $z$ variable. Field $\omega$ appears as zeroth order in $C$ solution to \eqref{Weq11}.

Master-field $W$ should belong to a specific functional class $\mathbf{C}^0$. 
This particular class $\mathbf{C}^0$ and others, namely $\mathbf{C}^1$ and $\mathbf{C}^2$, can be described via generating function. For the $\mathbf{C}^r$ class, where $r=0,1,2$ and reflects the power in $\theta$, the corresponding generating function looks as follows
\begin{equation}\label{gen_class}
\int \mathscr{D}\rho\int_0^1 d\mathcal{T}\, \frac{(1-\mathcal{T})^{1-r}}{\mathcal{T}^{1-r}}\exp\{i\mathcal{T}z_\alpha(y-B(\rho))^\alpha+i(1-\mathcal{T})y^\alpha A_\alpha(\rho)-i\mathcal{T}B_\alpha(\rho) A^\alpha(\rho)\}\,.
\end{equation}
The shorthand notation for integration is understood as follows
\begin{equation}\label{measure}
\int \mathscr{D}\rho:=\int d\rho_1\ldots \dr \rho_n\, \mu(\rho_1,\ldots,\rho_n)\,.
\end{equation}
Measure $\mu(\rho_1,\ldots,\rho_n)$ contains theta- and delta-function of $\rho$s which effectively makes the domain of integration a compact subset of $\mathbb{R}^n$. $\rho$-dependent functions $A$ and $B$ are linear combinations of the derivatives of $\omega$ and $C$ w.r.t. full (anti)holomorphic variable, i.e.
\begin{align}
&A_\alpha(t,p_1,\ldots,p_N|\rho)=A^t(\rho)t_\alpha+A^1(\rho)p_{1\alpha}+\ldots +A^N(\rho)p_{N\alpha}\,,\\
&B _\alpha(t,p_1,\ldots,p_N|\rho)=B^t(\rho)t_\alpha+B^1(\rho)p_{1\alpha}+\ldots +B^N(\rho)p_{N\alpha}\,
\end{align}
with
\begin{equation}
p_{\alpha}:=-i\frac{\partial}{\partial \mathsf{y}_{C}^\alpha}\,,\;\; t_\alpha:=-i\frac{\partial}{\partial \mathsf{y}_\go^\alpha}\,.
\end{equation}
Next, we will use {\it exponential representation}
\begin{equation}\label{exponential_form}
\omega(y|x) := \exp\{iy^\alpha t_\alpha\} \omega(\mathsf{y}_t|x) \big|_{\mathsf{y}_t=0}, \qquad C(y|x) := \exp\{iy^\alpha p_\alpha\} C(\mathsf{y}_C|x) \big|_{\mathsf{y}_C=0}\,.
\end{equation}
Thus the exponential in \eqref{gen_class} should be understood as follows
\begin{multline}
\exp\big\{i\mathcal{T} z_\alpha (y-B(t,p_1,\ldots,p_N|\rho))^\alpha+i(1-\mathcal{T})y^\alpha A_\alpha(t,p_1,\ldots,p_N|\rho)-\\
-i\mathcal{T} B_\alpha(t,p_1,\ldots,p_N|\rho)A^\alpha(t,p_1,\ldots,p_N|\rho)\big\}\go(\mathsf{y}_\go|x)C(\mathsf{y}_{C_1}|x)\ldots C(\mathsf{y}_{C_N}|x)\Big|_{\mathsf{y}_{\go, C_1,\ldots, C_N}=0}\,.
\end{multline}
To obtain a representative from a specific class from generating expression \eqref{gen_class} one takes derivatives w.r.t. $B$ and $A$ thus canceling potentially dangerous poles in $\mathcal{T}$. Rigidly defined by \eqref{LambdaDef} the master-field belongs to $\mathbf{C}^1$. Indeed, it can be easily brought to the form \eqref{gen_class}
\begin{equation}
\Lambda[C]=-i\theta^\alpha \frac{\partial}{\partial p^\alpha}\int_0^1 d\mathcal{T}\, \exp\big\{i\mathcal{T}z_\alpha(y-p)^\alpha\big\}C(\mathsf{y}_C|x)
\end{equation}
with $B=-p$ and $A=0$.

Below we demonstrate how one gets dynamics of the form \eqref{sch1form},\eqref{sch0form} from \eqref{dxWeq1} -\eqref{LambdaDef}. In zeroth order in $C$ \eqref{Weq11} is simply
\begin{equation}
\dr_z W+\mathcal{O}(C)=0\,.
\end{equation}
General solution is $z$-independent function $\omega(y|x)$. Plugging this solution to \eqref{dxWeq1} and \eqref{Ceq1} we obtain
\begin{equation}\label{linear_one}
\dr_x \go+\go\ast \go +\mathcal{O}(C)=0\,,
\end{equation}
\begin{multline}\label{linear_zero}
\dr_x C\ast \gamma=\dr_z\{\go,\Lambda\}_\ast+\mathcal{O}(C^2)=(\go\ast C- C\ast \pi[\go])\ast \gamma+\mathcal{O}(C^2)\;\; \Longleftrightarrow\\
\Longleftrightarrow\;\; \dr_x C=\go\ast C-C\ast \pi[\go]+\mathcal{O}(C^2)\,.
\end{multline}
Here $\pi$ is the automorphism of the star-product algebra \eqref{star_infty}
\begin{equation}\label{automorphism}
\pi\big[\Gamma(z,y)\big]=\Gamma(-z,-y)\,.
\end{equation}
To obtain nonlinear corrections one needs to solve \eqref{Weq11} in higher orders. In the linear approximation \eqref{Weq11} casts into
\begin{equation}\label{dzWwC}
\dr_zW_{\go C}+\dr_z W_{C\go}+\go\ast \Lambda+\Lambda\ast \go+\dr_x\Lambda\Big|_{\go C}+\dr_x\Lambda\Big|_{C\go}=0\,.
\end{equation}
To clarify the notation we emphasise that system \eqref{dxWeq1}-\eqref{LambdaDef} just like Vasiliev system \cite{Vasiliev:1999ba} is consistent even when fields take values in arbitrary associative algebra. For that reason one can consistently project \eqref{Weq11} and in particular \eqref{dzWwC} onto a specific ordering and solve equations for master-fields in the specific ordering. Notation like $\dr_x\Lambda\big|_{\go C}$ and $\dr_x\Lambda\big|_{C\go}$ should be understood as follows
\begin{equation}\label{dx_Lambda_goC}
\dr_x\Lambda\big|_{\go C}= -\int_0^1 d\mathcal{T}\, \mathcal{T} \theta^\alpha z_\alpha e^{i\mathcal{T}z_\alpha y^\alpha}\dr_x C(-\mathcal{T}z|x)\Big|_{\go C}= - \int_0^1 d\mathcal{T}\, \mathcal{T} \theta^\alpha z_\alpha e^{i\mathcal{T}z_\alpha y^\alpha}(\go\ast C)(-\mathcal{T}z|x)\,,
\end{equation}
\begin{equation}\label{dx_Lambda_Cgo}
\dr_x\Lambda\big|_{C \go}= -\int_0^1 d\mathcal{T}\, \mathcal{T} \theta^\alpha z_\alpha e^{i\mathcal{T}z_\alpha y^\alpha}\dr_x C(-\mathcal{T}z|x)\Big|_{C\go}=  \int_0^1 d\mathcal{T}\, \mathcal{T} \theta^\alpha z_\alpha e^{i\mathcal{T}z_\alpha y^\alpha}(C\ast\pi[\go])(-\mathcal{T}z|x)\,.
\end{equation}
Sign alteration is due to $\{\dr x^\mu,\theta^\alpha\}=0$.
Equation \eqref{dzWwC} is consistent in each ordering, namely $\go C$ and $C\go$, hence a particular solution can be obtained by virtue of Poincare lemma
\begin{equation}\label{explicit_sol}
W_{\go C}=-\hmt_0 (\go \ast \Lambda)\,,\;\; W_{C\go}=-\hmt_0 (\Lambda\ast \go)\,,
\end{equation}
where operator $\hmt_0$ is defined as
\begin{equation}\label{W_sol}
\hmt_0\big(f(z,y|\theta)\big):=z^\alpha \frac{\partial}{\partial \theta^\alpha}\int_0^1 \frac{dt}{t}\, f(tz,y|t\theta)\,.
\end{equation}
Terms $\hmt_0 \big(\dr_x\Lambda|_{\go C}\big)$ and $\hmt_0\big(\dr_x \Lambda|_{C\go}\big)$ vanish kinematically as they are proportional to $\theta^\alpha z_\alpha$. Plugging solutions \eqref{W_sol} to \eqref{dxWeq1} and \eqref{Ceq1} we obtain
\begin{equation}\label{COMS_Chiral}
\dr_x \go+\go\ast \go=\underbrace{-\dr_x W_{\go C} -\dr_x W_{C\go}-\go \ast W_{\go C}-\go \ast W_{C\go}-W_{\go C}\ast \go-W_{C\go}\ast \go}+\mathcal{O}(C^2)\, ,
\end{equation}
\begin{equation}\label{Cubic_Chiral}
\dr_x C \ast \gamma=(C\ast \pi[\go]-\go \ast C)\ast \gamma+\underbrace{\dr_z\big(W_{\go C} \ast \Lambda+W_{C\go}\ast \Lambda+\Lambda\ast W_{\go C}+\Lambda\ast W_{C\go}\big)}+\mathcal{O}(C^3)\,.
\end{equation}
Or using the explicit form of $W_{\go C}$ and $W_{C\go}$ \eqref{explicit_sol} dynamics in zero-form sector can be put into the form
\begin{equation}\label{Cubic_Chiral_expl}
\dr_x C\ast \gamma=\dr_z\{\go,\Lambda\}_\ast+\underbrace{\dr_z\big\{-\hmt_0\{\go,\Lambda\}_\ast,\Lambda\big\}_\ast}+\mathcal{O}(C^3)\,,
\end{equation}
which is more handy for future analysis.

Underbraced terms of \eqref{COMS_Chiral} and linear in $C$ part of \eqref{Cubic_Chiral} produce so-called central on-mass-shell theorem\footnote{More specifically r.h.s. of \eqref{COMS_Chiral} produces only holomorphic part of the central on-mass-shell theorem. To get full free Fronsdal dynamics one needs to add also antiholomorphic part. This is reachable eather within \cite{Vasiliev:1990en} or within the generalization of \eqref{dxWeq1}-\eqref{LambdaDef} developed in \cite{Didenko:2026nag}.} \cite{Vasiliev:1999ba} if one considers $\omega$ as  a perturbation over $AdS_4$ . Underbraced terms of \eqref{Cubic_Chiral} (similarly \eqref{Cubic_Chiral_expl}) have the form $\Upsilon(y|x)\ast \gamma$ which is due to remarkable {\it projective idenities} which hold for an arbitrary function $W\in \mathbf{C}^0$
\begin{equation}\label{Proj1}
\dr_z\left(W\ast \Lambda\right)=-\left(\int \dr^2 u\, \dr^2 v \,e^{iu_\alpha v^\alpha}\, W(z,y+u)C(y+v)\right)\Bigg|_{z=-y}\ast \gamma\,,
\end{equation}
\begin{equation}\label{Proj2}
\dr_z\left(\Lambda\ast W\right)=\left(\int \dr^2 u \, \dr^2 v\, e^{iu_\alpha v^\alpha}\, C(y+u) W(z,-y-v)\right)\Bigg|_{z=-y}\ast \gamma \,. 
\end{equation}
In general the r.h.s. of \eqref{Cubic_Chiral}, \eqref{Cubic_Chiral_expl} could be of the form
\begin{equation}\label{unwanted}
\dr_x C(y|x)\ast \gamma=F(y|x)\ast \gamma+\theta_\alpha\theta^\alpha G(z,y|x),
\end{equation}
which enforces to introduce additional constraints to dynamics, namely $G(z,y|x)=0$.

Projective identities emerge due to the specific form of $\Lambda$ \eqref{LambdaDef} however there are other functions from $\mathbf{C}^1$ that share projective properties. This particular freedom is utilized in section \ref{Burn_mango} where we relax rigid definition for $\Lambda$ \eqref{LambdaDef} leaving $\Lambda$ to be a solution to \eqref{Lambdaeq1} however within the class of projective one-forms in $\theta$ described in the next section.

\section{Projective one-forms $\mathbf{C}^{1,\mathfrak{P}}$}\label{projective_forms}
There is a subclass of the so-called projective one-forms in $\theta$ for which we use notation $\mathbf{C}^{1,\mathfrak{P}}$. The main feature of the functions from this class is the following
\begin{equation}
\dr_z\big(f(z,y)\ast \Gamma(z,y|\theta)\big)=\tilde{f}(y)\ast \gamma\;\;\; \forall f\in\mathbf{C}^0\,, \;\; \forall \Gamma\in \mathbf{C}^{1,\mathfrak{P}}\,
\end{equation}
and similarly
\begin{equation}
\dr_z\big(\Gamma(z,y|\theta)\ast g(z,y)\big)=\tilde{g}(y)\ast \gamma\;\;\; \forall g\in\mathbf{C}^0\,, \;\; \forall \Gamma\in \mathbf{C}^{1,\mathfrak{P}}\,
\end{equation}
for some purely $y$-dependent functions $\tilde{f}$ and $\tilde{g}$. From the above properties one easily deduces the following
\begin{equation}
\mathbf{C}^0\ast \mathbf{C}^{1,\mathfrak{P}}\subset \mathbf{C}^{1,\mathfrak{P}}\,,\;\; \mathbf{C}^{1, \mathfrak{P}} \ast \mathbf{C}^0\subset \mathbf{C}^{1, \mathfrak{P}}\,.
\end{equation}
This property rests on the specific form of the pre-exponential of a function from $\mathbf{C}^{1,\mathfrak{P}}$ (see \cite{Didenko:2026nag}), namely
\begin{equation}
\Gamma(z,y|\theta)=\int_0^1 d\mathcal{T}\, \mathcal{T}\, \underline{\theta^\alpha (z+A)_\alpha}\, \exp\big\{i\mathcal{T}z_\alpha(y-B)^\alpha+i(1-\mathcal{T})y^\alpha A_\alpha-i\mathcal{T} B_\alpha A^\alpha\dots\big\}\,.
\end{equation}

Such functions can be easily expressed in terms of a simple generating expression via shifted homotopy approach\footnote{Shifted homotopy approach even though not being general somehow is general enough for (anti)holomorphic system. It allows expressing all the spin-local vertices explicitly obtained in \cite{Didenko:2024zpd} in a concise form \cite{Korybut:2025vdn}.} \cite{Gelfond:2018vmi}
\begin{equation}\label{projective_1form}
\hmt_A \gamma(z,y-B)\,.
\end{equation}
The operator $\hmt_A$ generalizes Poincare lemma \eqref{W_sol} and acts as follows
\begin{equation}\label{shifted_homotopy}
\hmt_A f(z,y|\theta)=(z+A)^\alpha\int_0^1\dfrac{dt}{t}f(tz-(1-t)A,y|t\theta)\,.
\end{equation}
The action on $\gamma$ gives
\begin{equation}
\hmt_A \gamma=\theta^\alpha(z+A)_\alpha \int_0^1 dt\, t \, \exp\{itz_\alpha y^\alpha-i(1-t)y^\alpha A_\alpha\}\,.
\end{equation} 
A novel and comprehensive modification as compared to \cite{Korybut:2025vdn}
comes from an additional shift of $y$ variable, thus finally we have
\begin{equation}
\hmt_A \gamma(z,y-B)=\theta^\alpha(z+A)_\alpha \int_0^1 dt\, t \, \exp\{itz_\alpha (y-B)^\alpha-i(1-t)y^\alpha A_\alpha-itB^\alpha A_\alpha+iB^\alpha A_\alpha\}\,.
\end{equation} 
Parameters $A$ and $B$ may depend on $p_i$ and on various additional integration variables, i.e.
\begin{equation}
\int \mathscr{D}\rho \hmt_{A(p_1,\ldots,p_n|\rho)}\gamma(z,y-B(p_1,\ldots, p_n|\rho))\,.
\end{equation}
Such a form makes it easier to write down the corresponding projective identities (see Appendix A for technical derivation details)
\begin{equation}\label{Proj_R}
\; \; \dr_z\Big(f(z,y)\ast \hmt_A\gamma(z,y-B)\Big)=\Big[e^{-iy^\alpha B_\alpha} f(-y+A,y+B)\Big]\ast \gamma\,,
\end{equation}
\begin{equation}\label{Proj_L}
\dr_z\Big(\hmt_A\gamma(z,y-B)\ast f(z,y)\Big)=\gamma\ast \Big[e^{+iy^\alpha B_\alpha} f(y+A,y+B)\Big]\,.
\end{equation}

Within $\mathbf{C}^{1,\mathfrak{P}}$ there are $\dr_z$-exact forms which can be easily identified using  representation \eqref{projective_1form}. Since topology in $z$-space is trivial\footnote{For formulation of HS theory with nontrivial topology in $(Z,Y)$ space see \cite{DeFilippi:2019jqq}, \cite{Diaz:2024kpr}.} all exact forms are solutions to 
\begin{equation}
\dr_z \int \mathscr{D}\rho \hmt_{A(p_1,\ldots,p_n|\rho)}\gamma(z,y-B(p_1,\ldots, p_n|\rho))=0\,.
\end{equation}
Using resolution of unity for the shifted homotopy 
\begin{equation}\label{shift_unity}
\{\dr_z,\hmt_A\}=1-h_A\,,\;\;\; h_{A}\big(f(z,y|\theta)\big):=f(-A,y|0)
\end{equation}
one easily obtains the following equation on the measure hidden in $\mathscr{D}\rho$ (cf. \eqref{measure})
\begin{equation}\label{measure_constraint}
\theta_\alpha \theta^\alpha e^{iz_\alpha y^\alpha}\int \mathscr{D}\rho\,\exp\{-iz_\alpha B^\alpha(p_1,\ldots,p_n|\rho)\}=0\,.
\end{equation}
In what follows we confine ourselves mainly to projective forms linear in $C$, hence $A$ and $B$ depend on a single $p$, thus the generic form of such a form is
\begin{equation}
\int d\rho_1 \, d\rho_2 \, \mu(\rho_1,\rho_2)\, \hmt_{\rho_1 p}\gamma(z,y-\rho_2 p)\,.
\end{equation}
Here we wrote measure explicitly, since for linear in $C$ forms, constraint \eqref{measure_constraint} has a transparent interpretation. Since all exact form are in the kernel of $\dr_z$ we have
\begin{multline}
\dr_z \int d\rho_1 \, d\rho_2 \, \mu(\rho_1,\rho_2)\, \hmt_{\rho_1 p}\gamma(z,y-\rho_2 p)=\int d\rho_1 \, d\rho_2 \, \mu(\rho_1,\rho_2)\, \gamma(z,y-\rho_2 p)=\\
=\frac{1}{2} \theta_\alpha \theta^\alpha\, e^{iz_\alpha y^\alpha}\, \int d\rho_2\,\left(\int d\rho_1\, \mu(\rho_1,\rho_2)\right)\, e^{-i\rho_2 z_\alpha p^\alpha}=0\,.
\end{multline}
From this form we can easily extract a condition on measure $\mu(\rho_1,\rho_2)$, namely
\begin{equation}\label{condition}
\int d\rho_1\, \mu(\rho_1,\rho_2)=0\,,\;\;\; \forall \rho_2\,.
\end{equation}
Now we are in position to write down the generic expression for exact projective one-form in $\theta$ as a general solution to equation
\begin{equation}\label{Proj_eq}
\dr_z \varepsilon[C](z,y)=C\ast \int d\rho_1 \, d\rho_2 \, \mu(\rho_1,\rho_2)\, \hmt_{\rho_1 p}\gamma(z,y-\rho_2 p)
\end{equation}
which looks as follows
\begin{equation}\label{generic_form}
\varepsilon[C](z,y)=C\ast \int d\rho_1\, d\rho_2\, \mu(\rho_1,\rho_2)\, \hmt_p\hmt_{\rho_1 p}\gamma(z,y-\rho_2 p)+\varepsilon_0[C](y)
\end{equation}
with $\mu(\rho_1,\rho_2)$ satisfying \eqref{condition}. In general one might choose different from $\hmt_0$ (which after star-exchange relations \cite{Didenko:2018fgx} turned into $\hmt_p$) homotopy operator to find a particular solution of \eqref{Proj_eq} but all this freedom is governed by the choice of arbitrary purely $y$-dependent function $\varepsilon_0[C](y)$.

\section{Deformed generating for the (anti)holomorphic sector of HS theory}\label{Burn_mango}
In this section we present slightly different formulation of \eqref{dxWeq1}-\eqref{LambdaDef} utilizing knowledge about projective forms $\mathbf{C}^{1,\mathfrak{P}}$. To avoid any future possible confusion we suplement all the fields of the deformed version with primes.
\begin{align}
&\dr_{x} W^\prime(z,y|x)+W^\prime(z,y|x)*W^\prime(z,y|x)=0\,,\label{dxWeq1_prime}\\
&\dr_z W^\prime(z,y|x)+\{W^\prime(z,y|x),\Lambda^\prime(z,y|x)\}_{*}+\dr_x\Lambda^\prime(z,y|x)=0\,,\label{Weq11_prime}\\
&\dr_z\Lambda^\prime(z,y|x)=C^\prime(y|x)*\gga\,,\;\; \gamma:=\half\theta_\alpha\theta^\alpha e^{iz_\alpha y^\alpha}\,,\label{Lambdaeq1_prime}\\
&\dr_x C^\prime(y|x)*\gga=\dr_z\{W^\prime(z,y|x),\Lambda^\prime(z,y|x)\}_{*}\,,\label{Ceq1_prime}\\
&W^\prime(z,y|x)\in \mathbf{C}^0\, , \label{class1_prime}\\
&\Lambda^\prime(z,y|x) \in \mathbf{C}^{1,\mathfrak{P}}\label{LambdaDef_prime}
\end{align}
The only difference is that now $\Lambda$-field is not rigidly defined as \eqref{LambdaDef} but a solution to \eqref{Lambdaeq1_prime} within class of projective one-forms \eqref{LambdaDef_prime}.

Generic solution to \eqref{Lambdaeq1_prime} can be written as 
\begin{equation}\label{47}
\Lambda^\prime(z,y|x)=\int_0^1 d\mathcal{T}\, \mathcal{T} \theta^\alpha z_\alpha e^{i\mathcal{T}z_\alpha y^\alpha}C^\prime(-\mathcal{T}z|x)+\dr_z\varepsilon
\end{equation}
where $\dr_z\varepsilon\in \mathbf{C}^{1,\mathfrak{P}}$. Before we proceed some clarifications are in order. When it comes to system \eqref{dxWeq1}-\eqref{LambdaDef} there is also a different from \eqref{47} understanding of a gauge symmetry in $z$-space considered, for example, in \cite{Faliakhov:2026yum} and \cite{Didenko:2023vna}. In those papers $\Lambda$-field is changed under some $z$-dependent parameter, say $\xi(z,y)$, but this $\xi(z,y)$ is constrained so that gauge transformation of $\Lambda$ is of the form
\begin{equation}
\delta_\xi\Lambda[C]=\Lambda[\delta_\xi C]\,.
\end{equation}
Which means that rigid definition of $\Lambda$ \eqref{LambdaDef} is still preserved. Those symmetries should be understood as symmetries of the  resulting dynamics \eqref{sch0form},\eqref{sch1form} acting effectively on $\go$ and $C$ mapping one solution to another. We consider different type of gauge symmetry, similar to Vasiliev system where gauge transformation in $z$-space appear as homogeneous solutions to equations on the $S$-field, eq. \eqref{Lambdaeq1_prime} in our case. These gauge symmetries do affect the form of the resulting vertices in terms of $\go$ and $C$ and this effect is the main interest of the current paper investigated from the perspective of the system \cite{Didenko:2022qga}.
 
For simplicity we look for dynamics of the same order as \eqref{COMS_Chiral}, \eqref{Cubic_Chiral}. For the $\dr_z \varepsilon$ to produce any effect on these vertices $\varepsilon$ should be linear\footnote{$\varepsilon$ of higher orders of $C$, say $n>1$, induce field redefinitions of order $n+1$ (see \eqref{N+1}).} in $C^\prime$, i.e. of the form \eqref{generic_form}. We choose solution to \eqref{Lambdaeq1_prime} of the following form 
\begin{equation}\label{Lambda_prime}
\Lambda^\prime(z,y|x)=\Lambda[C^\prime]+\dr_z \varepsilon[C^\prime]\,.
\end{equation}

In the zeroth order in $C^\prime$ systems \eqref{dxWeq1}-\eqref{LambdaDef} and \eqref{dxWeq1_prime}-\eqref{LambdaDef_prime} are identical and thus \eqref{dxWeq1_prime}-\eqref{LambdaDef_prime} produce the same dyamics \eqref{linear_one},\eqref{linear_zero}. To get to the interaction vertices we need to solve \eqref{Weq11_prime} for $W^\prime_{\go^\prime C^\prime}$ and $W^\prime_{C^\prime \go^\prime}$. Corresponding equations are of the following
\begin{equation}\label{dzWwC_prime}
\dr_zW^\prime_{\go^\prime C^\prime}+\dr_z W^\prime_{C^\prime\go^\prime}+\go^\prime\ast \Lambda^\prime+\Lambda^\prime\ast \go^\prime+\dr_x\Lambda^\prime\big|_{\go^\prime C^\prime}+\dr_x\Lambda^\prime\big|_{C^\prime\go^\prime}=0\,.\\
\end{equation}
Plugging explicit form of $\Lambda^\prime$ \eqref{Lambda_prime} we get
\begin{align}
& \dr_z W^\prime_{\go^\prime C^\prime}+\omega^\prime\ast \Lambda[C^\prime]+\omega^\prime\ast \dr_z \varepsilon[C^\prime]+\dr_x\Lambda[C^\prime]\Big|_{\go^\prime C^\prime}+\dr_x\dr_z \varepsilon[C^\prime]\Big|_{\go^\prime C^\prime}=0\,,\\
& \dr_z W^\prime_{C^\prime \go^\prime}+\Lambda[C^\prime]\ast \go^\prime+\dr_z \varepsilon[C^\prime]\ast \go^\prime+\dr_x\Lambda[C^\prime]\big|_{C^\prime \go^\prime}+\dr_x\dr_z \varepsilon[C^\prime]\big|_{C^\prime\go^\prime }=0
\end{align}
Here $\dr_x \varepsilon$ part should be treated exactly like $\dr_x \Lambda$ in \eqref{dx_Lambda_goC},\eqref{dx_Lambda_Cgo}. Since $\omega^\prime$ is $z$-independent each equation can be put to a form
\begin{align}
& \dr_z W^\prime_{\go^\prime C^\prime}+\omega^\prime\ast \Lambda[C^\prime]+\dr_x\Lambda[C^\prime]\Big|_{\go^\prime C^\prime}-\dr_z\Big(\omega^\prime\ast \varepsilon[C^\prime]+\dr_x \varepsilon[C^\prime]\Big|_{\go^\prime C^\prime}\Big)=0\,,\label{eq1}\\
& \dr_z W^\prime_{C^\prime \go^\prime}+\Lambda[C^\prime]\ast \go^\prime+\dr_x\Lambda[C^\prime]\big|_{C^\prime \go^\prime}+\dr_z\Big( \varepsilon[C^\prime]\ast \go^\prime-\dr_x \varepsilon[C^\prime]\big|_{C^\prime\go^\prime }\Big)=0\label{eq2}\,.
\end{align} 
Particular solutions reads as
\begin{align}
& W^\prime_{\go^\prime C^\prime }=-\hmt_0\big(\go^\prime \ast \Lambda[C^\prime]\big)+\omega^\prime\ast \varepsilon[C^\prime]+\dr_x \varepsilon[C^\prime]\Big|_{\go^\prime C^\prime}\,,\label{W_prime_goC}\\
& W^\prime_{C^\prime\go^\prime}=-\hmt_0\big(\Lambda[C^\prime]\ast \go^\prime\big)-\varepsilon[C^\prime]\ast \go^\prime+\dr_x \varepsilon[C^\prime]\big|_{C^\prime\go^\prime }\, \label{W_prime_Cgo}.
\end{align}
Note that shift of $W$ driven by $\varepsilon$ has no effect on vertices in linear order, since this shift is just a gauge transformation. Indeed,
\begin{equation}\label{gauge_varepsilon}
W^\prime_{\go^\prime C^\prime }+W^\prime_{C^\prime\go^\prime}=-\hmt_0\big(\go^\prime \ast \Lambda[C^\prime]+\Lambda[C^\prime]\ast \go^\prime\big)+D_{\go^\prime}\varepsilon\,
\end{equation}
where $D_{\go^\prime}\varepsilon=\dr_x \varepsilon+\go^\prime\ast \varepsilon-\varepsilon\ast \go^\prime$ is the covariant exterior derivative with respect to $\go^\prime$. On the other hand vertices in the zero-forms acquire certain changes.

Consider \eqref{Ceq1_prime} 
\begin{equation}
\dr_x C^\prime\ast \gamma=\dr_z\Big\{W^\prime_{\go^\prime C^\prime }+W^\prime_{C^\prime \go^\prime},\Lambda[C^\prime]+\dr_z \varepsilon[C^\prime]\Big\}_\ast\,.
\end{equation}
Using explicit expressions for the master-fields   \eqref{Lambda_prime},\eqref{W_prime_goC}, \eqref{W_prime_Cgo} we can separate additional shift to the vertices driven by $\varepsilon$
\begin{multline}\label{Cubic_Chiral_prime}
\dr_x C^\prime\ast \gamma=\dr_z\Big\{-\hmt_0\{\go^\prime,\Lambda[C^\prime]\},\Lambda[C^\prime]\Big\}_\ast+\dr_z\Big\{-\hmt_0\{\go^\prime,\Lambda[C^\prime]\},\dr_z \varepsilon[C^\prime]\Big\}_\ast+\\+\dr_z\Big\{\dr_x \varepsilon[C^\prime]+\go^\prime\ast \varepsilon[C^\prime]-\varepsilon[C^\prime]\ast \go^\prime,\Lambda[C^\prime]+\dr_z \varepsilon[C^\prime]\Big\}_\ast\,.
\end{multline}
The first term on the r.h.s. is precisely the underbraced term of \eqref{Cubic_Chiral_expl} while the rest is deformation. The reasonable question is what field redefinition maps vertices \eqref{Cubic_Chiral_expl} \eqref{Cubic_Chiral_prime} into one another.

In general one can add purely $y$-dependent function, say $W^0_{\go C}(y)$ and $W^0_{C\go}(y)$, to \eqref{W_prime_goC} and \eqref{W_prime_Cgo} respectively to obtain different particular solution of \eqref{eq1}, \eqref{eq2}. We avoid such a modification because it would change the form of linear in $C$ vertices in one-forms, which are already identical in the framework of the Vasiliev system \cite{Didenko:2015cwv} and in the framework of the (anti)holomorphic system \cite{Didenko:2024zpd}. Thus no modification is required and thankfully $\varepsilon$-driven terms are pure gauge.

\section{Field redefinitions}\label{Field_redefinitions}
Before we proceed with the specific change of the field frame we consider the general case first. Suppose we have the following dynamics in zero-form sector
\begin{equation}\label{test_dyn}
\dr_x C=\Upsilon(\go,C)+\Upsilon(\go,C,C)+\ldots
\end{equation}
and we want to change the field frame as
\begin{equation}\label{change}
C\rightarrow C^\prime+f(C^\prime,C^\prime)\,.
\end{equation}
As a result we want dynamics on $C^\prime$ of the form
\begin{equation}\label{result_dyn}
\dr_x C^\prime=\Upsilon^\prime(\go,C^\prime)+\Upsilon^\prime(\go,C^\prime,C^\prime)+\ldots
\end{equation}
One straightforwardly substitute $C$ in \eqref{test_dyn} according to \eqref{change} 
\begin{equation}\label{formal}
\dr_x\big(C^\prime+f(C^\prime,C^\prime)\big)=\Upsilon\Big(\go,C^\prime+f(C^\prime,C^\prime)\Big)+\Upsilon\Big(\go,C^\prime+f(C^\prime,C^\prime),C^\prime+f(C^\prime,C^\prime)\Big)+\ldots
\end{equation}
While r.h.s. is already of the desired form, l.h.s., namely $\dr_x f(C^\prime,C^\prime)$, needs to be expressed in different terms.  Up to the second order in $C^\prime$ we have
\begin{equation}
\dr_x C^\prime=\Upsilon(\go,C^\prime)+\mathcal{O}(C^{\prime\, 2})\,.
\end{equation}
Thus we have
\begin{equation}\label{f_2}
\dr_x f(C^\prime,C^\prime)=f\Big(\Upsilon(\go,C^\prime),C^\prime\Big)+f\Big(C^\prime,\Upsilon(\go,C^\prime)\Big)+\mathcal{O}(C^{\prime\, 3})\,.
\end{equation}
Plugging \eqref{f_2} to \eqref{formal} we obtain \eqref{result_dyn} up to $\mathcal{O}(C^{\prime\, 3})$
\begin{multline}\label{result3}
\dr_x C^\prime=\Upsilon(\go,C^\prime)+\\
+\Upsilon\Big(\go,f(C^\prime,C^\prime)\Big)-f\Big(\Upsilon(\go,C^\prime),C^\prime\Big)-f\Big(C^\prime,\Upsilon(\go,C^\prime)\Big)+\Upsilon(\go,C^\prime,C^\prime)+\mathcal{O}(C^{\prime\, 3})\,.
\end{multline}
Primed vertex of the second order in $C^\prime$ thus looks as follows
\begin{equation}\label{vertex2}
\Upsilon^\prime(\go,C^\prime,C^\prime)=\Upsilon\Big(\go,f(C^\prime,C^\prime)\Big)-f\Big(\Upsilon(\go,C^\prime),C^\prime\Big)-f\Big(C^\prime,\Upsilon(\go,C^\prime)\Big)+\Upsilon(\go,C^\prime,C^\prime)
\end{equation}
In general one can proceed further finding \eqref{result_dyn} up to the forth order. For this one should use the second order vertices \eqref{vertex2} to compute \eqref{f_2}. 

Formula \eqref{vertex2} is useful when certain $f(C,C)$ is known already, in contrast we want to find $f(C,C)$. For that purpose it is better to have \eqref{vertex2} in the form 
\begin{equation}\label{vertex2_5}
\Upsilon^\prime(\go,C^\prime,C^\prime)=\Upsilon\Big(\go,f(C^\prime,C^\prime)\Big)+\Upsilon(\go,C^\prime,C^\prime)-\dr_x f(C^\prime,C^\prime)\,.
\end{equation}
Moreover, search for such $f(C^\prime,C^\prime)$ is simplified because of the fact that linear in $C$ vertices in both systems are identical. Additionally, dynamics in zero-form sector \eqref{Ceq1},\eqref{Ceq1_prime} always comes multiplied by $\gamma$, thus after multiplication and plugging the explicit expression one has
\begin{multline}\label{main}
\dr_z\Big\{-\hmt_0\{\go,\Lambda[C^\prime]\},\dr_z \varepsilon[C^\prime]\Big\}_\ast+\dr_z\Big\{\dr_x \varepsilon[C^\prime]+[\go,\varepsilon[C^\prime]],\Lambda[C^\prime]+\dr_z \varepsilon[C^\prime]\Big\}_\ast=\\
=\dr_z\Big\{\go,\Lambda\big[f(C^\prime,C^\prime)\big]\Big\}_\ast-\dr_x f(C^\prime,C^\prime)\ast \gamma\,.
\end{multline}
{In this form, the equation for $f(C^\prime,C^\prime)$ appears rather complicated. We will use various identities to rewrite the left‑hand side of \eqref{main} in a different way, which makes the equation for the unknown $f(C^\prime,C^\prime)$ simpler.} In what follows we omit writing primes and arguments of $\Lambda$ and $\varepsilon$ explicitly when it does not produce ambiguity. The first term on the l.h.s. of \eqref{main}
\begin{multline}\label{first}
\dr_z\Big\{-\hmt_0\{\go,\Lambda\}_\ast,\dr_z \varepsilon\Big\}_\ast=\dr_z\Big(-\hmt_0\{\go,\Lambda\}_\ast \ast \dr_z \varepsilon-\dr_z\varepsilon\ast \hmt_0\{\go,\Lambda\}_\ast\Big)=\\
=\dr_z\Big(\dr_z\big(\hmt_0\{\go,\Lambda\}_\ast \ast \varepsilon\big)-\dr_z\big(\hmt_0\{\go,\Lambda\}\big)\ast \varepsilon-\dr_z\big(\varepsilon\ast \hmt_0\{\go,\Lambda\}_\ast\big)+\varepsilon\ast \dr_z\hmt_0\{\go,\Lambda\}_\ast\Big)=\\
=\dr_z\Big[-\dr_z\hmt_0\{\go,\Lambda\}_\ast,\varepsilon\Big]_\ast\,.
\end{multline}
Using resolution of identity for $\hmt_0$ (cf. \eqref{shift_unity}), namely
\begin{equation}\label{unity}
\{\dr_z,\hmt_0\}=1-h_0\,, \;\;\; h_0\big(f(z,y|\theta)\big)=f(0,y|0)\,,
\end{equation}
expression \eqref{first} can be simplified futher to the form
\begin{equation}
\dr_z\Big\{-\hmt_0\{\go,\Lambda\}_\ast,\dr_z \varepsilon\Big\}_\ast=\dr_z\Big[-\{\go,\Lambda\}_\ast,\varepsilon\Big]_\ast+\dr_z\Big[\hmt_0\dr_z\{\go,\Lambda\}_\ast,\varepsilon\Big]_\ast\,.
\end{equation}
According to linear dynamics \eqref{linear_zero} $\dr_z\{\go,\Lambda\}_\ast$ equals to $\dr_x C\ast \gamma$, thus previous expression finally acquires the form
\begin{equation}
\dr_z\Big\{-\hmt_0\{\go,\Lambda\}_\ast,\dr_z \varepsilon\Big\}_\ast=\dr_z\Big[-\{\go,\Lambda\}_\ast,\varepsilon\Big]_\ast+\dr_z\Big[-\dr_x\Lambda,\varepsilon\Big]_\ast\,.
\end{equation}
In a similar fashion one can transform the second term of the l.h.s. of \eqref{main}. Before we proceed with the final answer, we would like to note where $z$-indepence of $\omega$ matters. Consider 
\begin{multline}
\dr_z\big\{[\go,\varepsilon]_\ast,\dr_z\varepsilon\big\}_\ast=\dr_z\big(\go\ast \varepsilon\ast \dr_z\varepsilon-\varepsilon\ast\go\ast \dr_z \varepsilon+\dr_z \varepsilon\ast \go\ast\varepsilon-\dr_z\varepsilon \ast \varepsilon\ast\go\big)=\\
=\dr_z\big(\go\ast \varepsilon\ast \dr_z\varepsilon+\dr_z (\varepsilon\ast \go\ast\varepsilon)-\dr_z\varepsilon \ast \varepsilon\ast\go\big)=\dr_z\big\{\go,\varepsilon\ast \dr_z\varepsilon\big\}_\ast\,.
\end{multline}
Eventually l.h.s. of \eqref{main} is of the form
\begin{multline}
\dr_z\big\{-\hmt_0\{\go,\Lambda\},\dr_z \varepsilon\big\}_\ast+\dr_z\big\{\dr_x \varepsilon+[\go,\varepsilon],\Lambda+\dr_z \varepsilon\big\}_\ast=\\
=\dr_z\big\{\go,[\varepsilon,\Lambda]_\ast+\varepsilon\ast \dr_z\varepsilon\big\}_\ast+\dr_z\dr_x\big([\varepsilon,\Lambda]_\ast+\varepsilon\ast \dr_z\varepsilon\big)\,.
\end{multline}
This particular form suggests that field redefinition should be of the form
\begin{equation}\label{Batyaev}
f(C,C)\ast \gamma=\dr_z\big([\varepsilon,\Lambda]_\ast+\varepsilon\ast \dr_z\varepsilon\big)\,.
\end{equation}
To obtain $\Lambda\big[f(C,C)\big]$ we observe that
\begin{equation}
\Lambda\big[f(C,C)\big]=\hmt_0\big(f(C,C)\ast\gamma\big)=\hmt_0\dr_z \big([\varepsilon,\Lambda]_\ast+\varepsilon\ast \dr_z\varepsilon\big)\,.
\end{equation}
Plugging this expression to \eqref{main} with the of resolution of identity \eqref{unity} and using $z$-independence of $\omega$ one easily consludes that \eqref{main} for $f(C,C)$ given by \eqref{Batyaev} holds. Note that each term, namely $[\varepsilon,\Lambda]_\ast$ and $\varepsilon\ast \dr_z\varepsilon$ belong to $\mathbf{C}^{1,\mathfrak{P}}$, thus $f(C,C)$ is only $y$-dependent.

Analogously to \eqref{Lambda_prime} one can deform $\Lambda$ with $\varepsilon[C,\ldots,C]$ of the $n$-th order for $n>1$. In this case vertices up to $n$-th order are the same for \eqref{dxWeq1}-\eqref{LambdaDef} and \eqref{dxWeq1_prime}-\eqref{LambdaDef_prime}, however in the $n+1$ order zero-form vertices are related by the change of the field frame of the form
\begin{equation}\label{N+1}
f(\underbrace{C,\ldots,C}_{n+1})\ast \gamma=\dr_z\big[\varepsilon[\underbrace{C,\ldots,C}_n],\Lambda[C]\big]_\ast\,.
\end{equation}
Here $\varepsilon\ast\dr_z \varepsilon$ term is absent because it exceeds the order  of $n+1$ in $C$. 

\section{Explicit form of $f[C,C]$ and its nontriviality}\label{Explicit_f}
Using explicit form for linear exact projective form \eqref{generic_form} we can compute \eqref{Batyaev}. We denote $z$-dependent part of $\varepsilon$ as $\varepsilon_1$
\begin{equation}\label{varepsilon_1}
\varepsilon_1[C](z,y):=C\ast \int d\rho_1\, d\rho_2\, \mu(\rho_1,\rho_2)\, \hmt_p\hmt_{\rho_1 p}\gamma(z,y-\rho_2 p)\,.
\end{equation} 
Field redefinition \eqref{Batyaev} appears as a sum of terms driven by $\varepsilon_0$ and $\varepsilon_1$
\begin{equation}
f(C,C)\ast \gamma=\dr_z[\varepsilon_0,\Lambda]_\ast+\dr_z\big([\varepsilon_1,\Lambda]_\ast+\varepsilon_1\ast \dr_z \varepsilon_1\big)
\end{equation}
where the following expression is used due to z-independence of $\varepsilon_0(y)$
\begin{equation}
\mathrm{d}_z\big((\varepsilon_1(z,y)+\varepsilon_0(y))*\mathrm{d}_z \varepsilon_1(z,y)     \big)=\mathrm{d}_z\big(\varepsilon_1(z,y)*\mathrm{d}_z \varepsilon_1(z,y)\big)
\end{equation}
The contribution from $z$-independent part $\varepsilon_0[C](y)$ is pure gauge, i.e.
\begin{equation}\label{twisted_commutator}
f_0[C,C]\ast \gamma=\big(\varepsilon_0\ast C-C\ast \pi[\varepsilon_0]\big)\ast \gamma\,,
\end{equation}
and thus has no effect on vertices. While contribution driven by $\varepsilon_1$
\begin{equation}
f_1[C,C]\ast \gamma=\dr_z\big([\varepsilon_1,\Lambda]_\ast+\varepsilon_1\ast \dr_z \varepsilon_1\big)
\end{equation}
in nontrivial as we demonstrate below. Each contribution in terms of $\mu(\rho_1,\rho_2)$ satisfying \eqref{condition} is of the form
\begin{multline}\label{e_Lambda}
\dr_z(\varepsilon_1\ast \Lambda)=\int \mu(\rho_1,\rho_2)\int_0^1 d\mathcal{T}\, (1-\mathcal{T})\int_0^1 d\sigma\, (1-\rho_1)\, (y^\alpha p_{1\alpha})\times\\
\times \exp\Big\{i\big(\mathcal{T}(1-\rho_2)+(1-\mathcal{T})\sigma (1-\rho_1)\big)y^\alpha p_{1\alpha}+i(1-\mathcal{T})y^\alpha p_{2\alpha}-\\
-i(1-\mathcal{T})\sigma (1-\rho_1)p_2 {}^\alpha p_{1\alpha}\Big\}C(\mathsf{y}_1|x)C(\mathsf{y}_2|x)\Big|_{\mathsf{y}_{1,2}=0}\ast \gamma\,,
\end{multline}

\begin{multline}\label{Lambda_e}
\dr_z (\Lambda\ast \varepsilon_1)=\int \mu(\rho_1,\rho_2) \int_0^1 d\mathcal{T}\, (1-\mathcal{T})\int_0^1 d\sigma\, (1-\rho_1)\, (y^\alpha p_{2\alpha})\times\\
\times \exp\Big\{i(1-\mathcal{T})y^\alpha p_{1\alpha}+i\big(\mathcal{T}(1-\rho_2)-(1-\mathcal{T})\sigma (1-\rho_1)\big)y^\alpha p_{2\alpha}+\\+i(1-\mathcal{T})\sigma(1-\rho_1)p_2 {}^\alpha p_{1\alpha}\Big\}C(\mathsf{y}_1|x)C(\mathsf{y}_2|x)\Big|_{\mathsf{y}_{1,2}=0}\ast \gamma\,,
\end{multline}
\begin{multline}\label{e_de}
\dr_z(\varepsilon_1\ast \dr_z \varepsilon_1)=\int \mu(\tilde{\rho}_1,\tilde{\rho}_2)\int \mu(\rho_1,\rho_2)\int_0^1 d\mathcal{T}\, (1-\mathcal{T}) \int_0^1 d\sigma\, (1-\rho_1)\Big(y^\alpha p_{1\alpha}+(1-\tilde{\rho}_1)p_2 {}^\alpha p_{1\alpha}\Big)\times\\
\times\exp\Big\{i\big(\mathcal{T}(1-\rho_2)+(1-\mathcal{T})\sigma(1-\rho_1)\big)y^\alpha p_{1\alpha}+i\big((1-\mathcal{T})(1-\tilde{\rho}_2)-\mathcal{T}(1-\tilde{\rho}_1)\big)y^\alpha p_{2\alpha}+\\
+i\big(\mathcal{T}(1-\tilde{\rho}_1)(1-\rho_2)-(1-\mathcal{T})\sigma(1-\rho_1)(1-\tilde{\rho}_2)\big)p_2 {}^\alpha p_{1\alpha}\Big\}C(\mathsf{y}_1|x)C(\mathsf{y}_2|x)\Big|_{\mathsf{y}_{1,2}=0}\ast \gamma\,,
\end{multline}

To analyze corresponding contribution we are going to use {\it exponential} representation, i.e. we consider functions of $y,p_1,p_2$ acting on $C(\mathsf{y}_1|x)C(\mathsf{y}_2|x)\big|_{\mathsf{y}_{1,2}=0}$. Since Lorentz invariance is preserved $y,p_1,p_2$ appear only as contractions
\begin{equation}\label{XYZ}
X:=y^\alpha p_{1\alpha}\,,\;\; Y:=y^\alpha p_{2\alpha}\,,\;\; Z:=p_{1\alpha} p_2 {}^\alpha\,.
\end{equation}
Note there are only 3 spinors hence no Schouten identities are available thus $X^m Y^n Z^k$ is the basis. 

For field redefinition driven by $\varepsilon_0$ and  $\varepsilon_1$ we have respectively
\begin{equation}
f_0[C,C]=\big(\varepsilon_0\ast C-C\ast \pi[\varepsilon_0]\big)=\mathscr{F}_0(X,Y.Z)C(\mathsf{y}_1|x)C(\mathsf{y}_2|x)\big|_{\mathsf{y}_{1,2}=0}\,,
\end{equation}
\begin{equation}
f_1[C,C]\ast \gamma=\dr_z\big([\varepsilon_1,\Lambda]_\ast +\varepsilon_1\ast \dr_z\varepsilon_1\big)=\mathscr{F}_1(X,Y,Z)C(\mathsf{y}_1|x)C(\mathsf{y}_2|x)\big|_{\mathsf{y}_{1,2}=0}\ast \gamma\,.
\end{equation}
The function $\mathscr{F}_0(X,Y,Z)$ obey several constraints, namely (see Appendix B for the proof)
\begin{equation}\label{cyclic1}
\mathscr{F}_0(X,Y,Z)+\mathscr{F}_0(Y,Z,X)+\mathscr{F}_0(Z,X,Y)=0\,,
\end{equation}
\begin{equation}\label{factor1}
\partial_X \partial_Y\Big(e^{-i(X+Y)}\mathscr{F}_0(X,Y,Z)\Big)=0\,,
\end{equation}
\begin{equation}\label{strange1}
\big(\partial_X+\partial_Y+\partial_Z\big)\mathscr{F}_0(X,Y,Z)=i\mathscr{F}_0(X,Y,Z)\,.
\end{equation}
These constraints are necessary and sufficient, i.e. for the function $\mathscr{F}_0(X,Y,Z)$ obeying \eqref{cyclic1}-\eqref{strange1} one can always find corresponding $\varepsilon_0(y)$ such that $f_0[C,C]$ is the twisted commutator \eqref{twisted_commutator}. Indeed, consider
\begin{equation}
\Phi(X-Z)=\frac{i}{2} e^{-iY}\Big(e^{iX}\partial_X\big(e^{-iX}\mathscr{F}(X,Y,Z)\big)-e^{iZ}\partial_Z\big(e^{-iZ} \mathscr{F}(Y,Z,X)\big)\Big)\,.
\end{equation}
Here r.h.s. does depend only on $(Z-Y)$ provided \eqref{cyclic1}-\eqref{strange1} are fulfilled. $\varepsilon_0$ is then of the form
\begin{equation}
\varepsilon_0[C](y)=\Phi(y^\alpha p_\alpha)C(\mathsf{y}|x)\big|_{\mathsf{y}=0}\,.
\end{equation}

To show that $\varepsilon_1$ provides a nontrivial deformation we analyze $\mathscr{F}_1(X,Y,Z)$ against conditions \eqref{cyclic1}-\eqref{strange1}. Function $\mathscr{F}_1(X,Y,Z)$ is analytic in its arguments as seen from \eqref{e_Lambda}-\eqref{e_de} thus it can be written as
\begin{equation}
\mathscr{F}_1(X,Y,Z)=\sum_{m,n,k} C_{m,n,k} X^mY^n Z^k\,.
\end{equation}
Moreover, those expansion coefficients are determined by various moments of measure $\mu(\rho_1,\rho_2)$
\begin{equation}\label{moments}
\mu_{k,l}:=\int d\rho_1\, d\rho_2\, \mu(\rho_1,\rho_2)\, (1-\rho_1)^k(1-\rho_2)^l\,
\end{equation}
and constraint \eqref{condition} from momenta perspective looks as 
\begin{equation}\label{condition_moment}
\mu_{0,k}=0\;\;\forall k\in\mathbb{N}_0\,.
\end{equation}
Already in the lowest orders one can easily see that the cyclic constraint \eqref{cyclic1} fails, i.e.
\begin{equation}
C_{m,n,k}+C_{n,k,m}+C_{k,m,n}\neq 0\,.
\end{equation}
For example
\begin{equation}\label{C111}
C_{1,1,1}=\frac{1}{4}\mu_{2,0}-\frac{1}{6}\mu_{1,1}^2+\frac{1}{24}\mu_{2,0}^2\,.
\end{equation}
Thus in general $\varepsilon_1$-deformation cannot be gauged away.

We are not yet in position to present the full list of constraints for $\mathscr{F}_1(X,Y,Z)$ like the one presented for $\mathscr{F}_0(X,Y,Z)$. The main technical complexity emerges from its nonlinear nature, namely \eqref{e_de} part. This problem will be analyzed elsewhere. Nonetheless specific choices of $\mu(\rho_1,\rho_2)$ lead to local field redefinitions. In terms of expansion coefficients locality means
\begin{equation}\label{locality_constraint}
C_{m,n,k}=0\;\;\; \forall k >N\,, \forall m,n \geq 0\,.
\end{equation}
for some $N\in \mathbb{N}_0$.

\subsection*{Local regime}
In general measure $\mu(\rho_1,\rho_2)$ is allowed to be a distribution. Roughly speaking one get any set of moments $\{\mu_{m,n}\}$ of such measure \eqref{moments} choosing the measure of the form
\begin{equation}\label{distribution}
\mu(\rho_1,\rho_2)=\sum_{m=0}^\infty \sum_{n=0}^\infty \mu_{m,n}\, m!\, n!\,(-1)^{m+n}\delta^{(m)}(1-\rho_1)\delta^{(n)}(1-\rho_2)\,.
\end{equation}
Here $\delta^{(m)}(1-\rho_1)$, $\delta^{(n)}(1-\rho_2)$ are $m$th and $n$th derivatives w.r.t. the full argument. Imposing constraint \eqref{locality_constraint} one finds that it can be fulfilled properly adjusting higher and higher moments. For example, $C_{1,1,2}$ compared to \eqref{C111} is expressed in terms higher moment
\begin{equation}
C_{1,1,2}=-\frac{i}{120}(7\mu_{2,1}^2-18\mu_{1,2}\mu_{3,0})\,.
\end{equation}  
However for \eqref{distribution} to converge in the space of distribution moments should decay rapidly. Analysis of the decay rate of the moments in the most general setup is beyond the scope of the current paper. Nonetheless hard cutoff works just fine. One of the simplest ways to obtain local $f_1[C,C]$ is by choosing $\mu(\rho_1,\rho_2)$ of the form
\begin{equation}\label{local_measure}
\mu^{loc}(\rho_1,\rho_2)=-\delta^\prime(1-\rho_1)\widetilde{\mu}(\rho_2)\,.
\end{equation}
Here $\widetilde{\mu}(\rho_2)$ is arbitrary integrable function. Such measure obviously  bypasses constraint \eqref{condition} and after simple partial integration w.r.t. $\rho_1$ contributions \eqref{e_Lambda}-\eqref{e_de} acquire the following form respectively
\begin{equation}
\mathscr{F}_1^{\dr_z(\varepsilon_1\ast\Lambda)}(X,Y,Z)=\int d\rho_2\,  \widetilde{\mu}(\rho_2)\int_0^1d\mathcal{T}\, (1-\mathcal{T})\, X\, e^{i\mathcal{T}(1-\rho_2)X+i(1-\mathcal{T})Y},
\end{equation} 
\begin{equation}
\mathscr{F}_1^{\dr_z(\Lambda\ast \varepsilon_1)}(X,Y,Z)=\int d\rho_2 \, \widetilde{\mu}(\rho_2)\int_0^1d\mathcal{T}\,(1-\mathcal{T})\,Y\, e^{i(1-\mathcal{T})X+i\mathcal{T}(1-\rho_2)Y}\,,
\end{equation}
\begin{multline}
\mathscr{F}_1^{\dr_z(\varepsilon_1\ast\dr_z\varepsilon_1)}(X,Y,Z)=\int d\rho_2 \, \widetilde{\mu}(\rho_2)\int d\tilde{\rho}_2\,   \widetilde{\mu}(\tilde{\rho}_2)\int_0^1 d\mathcal{T}\,(1-\mathcal{T})\times\\
\times\Big[X\big(i\mathcal{T}(1-\rho_2)Z-i\mathcal{T}Y\big)+Z\Big]e^{i\mathcal{T}(1-\rho_2)X+i(1-\mathcal{T})(1-\tilde{\rho}_2)Y}\,,
\end{multline}
where it is similar $X:=y^\alpha p_{1\alpha}\,,\;\; Y:=y^\alpha p_{2\alpha}\,,\;\; Z:=p_{1\alpha} p_2 {}^\alpha\,$. Full $\mathscr{F}_1(X,Y,Z)$ driven by $\varepsilon_1$ with local measure \eqref{local_measure} is then
\begin{equation}
\mathscr{F}^{loc}_1(X,Y,Z)=\mathscr{F}_1^{\dr_z(\varepsilon_1\ast\Lambda)}(X,Y,Z)-\mathscr{F}_1^{\dr_z(\Lambda\ast \varepsilon_1)}(X,Y,Z)+\mathscr{F}_1^{\dr_z(\varepsilon_1\ast\dr_z\varepsilon_1)}(X,Y,Z)\,.
\end{equation}

In general one can consider other versions for local measure
\begin{equation}
\mu(\rho_1,\rho_2)=\delta^{(2)}(1-\rho_1)\widetilde{\mu}_2(\rho_2)+\delta^{(3)}(1-\rho_1)\widetilde{\mu}_3(\rho_2)+\ldots
\end{equation}
with hard cutoff at some order. Those measures also lead to local field redefinition however with more derivatives, i.e. powers of $Z$ \eqref{XYZ}.
 
\section{Conclusion}\label{Conclusion}

A consistent deformation of the generating system proposed in \cite{Didenko:2022qga} is considered. Our deformation relaxes the requirement for the master-field $\Lambda$ to be a fixed functional of the 
$C$-field. We allow $\Lambda$ to be a solution to the corresponding equation within the class of the so-called projective one-forms, which we denote as 
$\mathbf{C}^{1,\mathfrak{P}}$. This class was described in full generality in terms of a simple generating expression \eqref{projective_1form} useful for practical computations. Belonging to this class guarantees the absence of unwanted constraints (cf. \eqref{unwanted}) due to the projective identities \eqref{Proj_R}, \eqref{Proj_L}, which were found here. We study a solution to the equation for $\Lambda$ \eqref{Lambdaeq1_prime}, which differs from the previous one by a shift with an exact form $\dr_z\varepsilon \in \mathbf{C}^{1,\mathfrak{P}}$ linear in $C$. Such a deformation, compared to the original formulation, induces a field redefinition, which was found explicitly in \eqref{Batyaev} for arbitrary admissible $\varepsilon$.

Our approach to $\mathbf{C}^{1,\mathfrak{P}}$ class allowed a quite explicit description of $\varepsilon\in\mathbf{C}^0$ such that $\dr_z \varepsilon\in \mathbf{C}^{1,\mathfrak{P}}$. In terms of integral representations, a simple integral constraint on the corresponding measure $\mu(\rho_1,\rho_2)$ \eqref{condition} suffices, which is equivalent to the vanishing of certain moments \eqref{condition_moment}. We presented the induced change of the field frame in terms of $\mu(\rho_1,\rho_2)$ \eqref{e_Lambda}--\eqref{e_de}, which completely governs all degrees of freedom of the deformation we introduce. Moreover, for specific choices of  $\mu(\rho_1,\rho_2)$, the corresponding field redefinitions admit a local form.

As anticipated, certain $\varepsilon$ do not produce an effect on vertices, inducing a gauge transformation instead. Those parameters are purely $y$-dependent functions $\varepsilon_0(y)\in\mathbf{C}^0$. We identified necessary and sufficient conditions \eqref{cyclic1}-\eqref{strange1} for the change of a field frame that corresponds to a pure gauge. We thus conclude that the $z$-dependent part of $\varepsilon\in\mathbf{C}^0$, namely $\varepsilon_1$ \eqref{varepsilon_1}, induces a nontrivial deformation which cannot be gauged away.
  
There is a famous change of a field frame originally introduced in \cite{Vasiliev:2016xui}
\begin{equation}\label{Vas_shift}
f^{Vas.}[C,C]=\frac{\eta}{2}\int d^3\tau\, \theta(\tau_1)\theta(\tau_2)\theta(\tau_3)\delta^\prime(1-\sum_{i=1}^3 \tau_i)\, e^{-i\tau_2 p_{1\alpha}p_2 {}^\alpha}C(\tau_1y)C(-\tau_3y)k\,
\end{equation}
that relates the nonlocal vertex obtained by virtue of the pure Poincaré lemma from the Vasiliev generating system to the spin-local one\footnote{An additional outer Klein operator $k$ is present due to the enlarged spectrum of the full Vasiliev system compared to the pure (anti)holomorphic system of \cite{Didenko:2022qga}. The Vasiliev system also describes so‑called topological fields \cite{Kirakosiants:2026spd}. The outer Klein operator can be consistently introduced in the (anti)holomorphic system, since \eqref{automorphism} is indeed an automorphism for the product \eqref{star_infty}.}. In the framework of the system \eqref{dxWeq1}-\eqref{LambdaDef} of \cite{Didenko:2022qga}, one obtains the same spin-local vertex again by virtue of the Poincaré lemma, but straightaway without introducing any additional field redefinitions. A yet unresolved question is whether one is able to obtain \eqref{Vas_shift} as a field redefinition induced from the deformation introduced here. The technical complexity of this analysis emerges from the nonlinearity of the corresponding field redefinition in terms of $\varepsilon$.

Moreover, it is worth mentioning that the effect of the shift of $\Lambda$ by $\dr_z \varepsilon[C]\in\mathbf{C}^{1,\mathfrak{P}}$ does not end in the second order. Even the linear deformation induces a change of the field in all orders, affecting both $\go$ and $C$ starting from the third order.

One can also consider a deformation of $\Lambda$ by $\dr_z\varepsilon\in \mathbf{C}^{1,\mathfrak{P}}$ which is of the $n$th order in 
$C$. Such a deformation induces a change of the field frame of the order 
$n+1$ \eqref{N+1} and beyond.

\section*{Acknowledgement}
We would like to thank Slava Didenko for fruitful discussions and valuable comments on the manuscript. This research was supported by the Russian Science Foundation grant 26-12-00240, https://rscf.ru/en/project/26-12-00240/.

\newcounter{appendix}
\setcounter{appendix}{1}
\renewcommand{\theequation}{\Alph{appendix}.\arabic{equation}}
\addtocounter{section}{1} \setcounter{equation}{0}
 \renewcommand{\thesection}{\Alph{appendix}.}

\addcontentsline{toc}{section}{\,\,\,\,\, Appendix A}

\section*{Appendix A}\label{AppA}

Projective identities \eqref{Proj_R},\eqref{Proj_L} for a generic $\mathbf{C}^{1,\mathfrak{P}}$ allow for robust computation. Below we consider one of them, namely
\begin{equation}\label{Proj_R_A}
\dr_z\Big(f(z,y)\ast \hmt_A\gamma(z,y-B)\Big)=\Big[e^{-iy^\alpha B_\alpha} f(-y+A,y+B)\Big]\ast \gamma
\end{equation}

According to \eqref{gen_class} generic element of $\mathbf{C}^0$ can be obtained from a generating function 
\begin{equation}
f(z,y)=\int \mathscr{D}\rho \int_0^1d\mathcal{T}\, \frac{1-\mathcal{T}}{\mathcal{T}}\, e^{i\mathcal{T}z_\alpha (y-\tilde{B})^\alpha +i(1-\mathcal{T})y^\alpha \tilde{A}_\alpha-i\mathcal{T}\tilde{B}_\alpha \tilde{A}^\alpha}\,
\end{equation}
with some $\rho$-dependent $\tilde{A}$ and $\tilde{B}$. Particular form this dependence and domain of integration of $\rho$s is unimportant, thus we suppress $\int \mathscr{D}\rho$ in what follows for brevity.  Recall that particular element of $\mathbf{C}^0$ is obtained by taking derivatives w.r.t. $\tilde{B}$ and $\tilde{A}$. Such derivatives bring additional power of $\mathcal{T}$ thus canceling potentially dangerous pole. 

Generating function for $\mathbf{C}^{1,\mathfrak{P}}$ looks as follows
\begin{equation}
\hmt_A \gamma(z,y-B)=\theta^\alpha(z+A)_\alpha \int_0^1 d\mathcal{T}\, \mathcal{T} \, \exp\{i\mathcal{T}z_\alpha (y-B)^\alpha-i(1-\mathcal{T})y^\alpha A_\alpha-i\mathcal{T}B^\alpha A_\alpha+iB^\alpha A_\alpha\}\,.
\end{equation} 
Computing star-product \eqref{star_infty} one obtains
\begin{multline}
f(z,y)\ast \hmt_A\gamma(z,y-B)=\\
=\int_0^1d\mathcal{T}\,\mathcal{T}\int_0^1d\sigma\,\frac{1-\sigma}{\sigma}\theta^\alpha (z+A-\tilde{A})_\alpha\, \exp\Big\{i\mathcal{T}z_\alpha\big(y-\sigma(\tilde{B}-A)+(1-\sigma)(B-\tilde{A})\big)^\alpha-i\mathcal{T}\sigma \tilde{B}_\alpha(\tilde{A}-A)^\alpha+\\
+i(1-\mathcal{T})\big(y^\alpha(\tilde{A}-A)_\alpha+\tilde{A}_\alpha A^\alpha\big)-i\mathcal{T}(1-\sigma)B_\alpha(\tilde{A}-A)^\alpha+iB^\alpha A_\alpha\Big\}\,.
\end{multline}
Applying $\dr_z$ to the previous one easily finds that integrand of the resulting expression turns into total derivative w.r.t. $\mathcal{T}$
\begin{equation}
\dr_z\big(f(z,y)\ast \hmt_A\gamma(z,y-B)\big)=\frac{1}{2}\theta_\beta \theta^\beta\int_0^1 d\sigma\, \frac{1-\sigma}{\sigma}\int_0^1d\mathcal{T}\,\big(2\mathcal{T}+\mathcal{T}^2\frac{\partial}{\partial \mathcal{T}}\big)\exp\{\ldots\}\,.
\end{equation}
After a simple partial integration it casts into
\begin{multline}
\dr_z\big(f(z,y)\ast \hmt_A\gamma(z,y-B)\big)=\frac{1}{2}\theta_\beta \theta^\beta \int_0^1 d\sigma\, \frac{1-\sigma}{\sigma}\, \exp\big\{iz_\alpha\big(y-\sigma(\tilde{B}-A)-(1-\sigma)(B-\tilde{A})\big)^\alpha+\\+i\sigma \tilde{B}_\alpha(A-\tilde{A})^\alpha+i(1-\sigma)B_\alpha(A-\tilde{A})^\alpha+iB^\alpha A_\alpha\big\}\,.
\end{multline}
The only $z$-dependent part of the l.h.s. is $e^{iz_\alpha y^\alpha}$. Hence by virtue of a simple identity
\begin{equation}
g(-y)\ast e^{iz_\alpha y^\alpha}=g(z)e^{iz_\alpha y^\alpha}
\end{equation}
one shows that \eqref{Proj_R_A} holds. 

In a similar fashion one shows that analogous identity, namely
\begin{equation}\label{Proj_L_A}
\dr_z\Big(\hmt_A\gamma(z,y-B)\ast f(z,y)\Big)=\gamma\ast \Big[e^{+iy^\alpha B_\alpha} f(y+A,y+B)\Big]\,,
\end{equation}
is also true.

\renewcommand{\theequation}{\Alph{appendix}.\arabic{equation}}
\addtocounter{appendix}{1} \setcounter{equation}{0}
\addtocounter{section}{1}
\addcontentsline{toc}{section}{\,\,\,\,\, Appendix B}

\section*{Appendix B}\label{AppB}

Necessary and sufficient conditions for the change of field frame to be a gauge transformation, namely to be of the form
\begin{equation}\label{gauge_0}
\varepsilon_0[C] \ast C-C\ast \pi\big[\varepsilon_0[C]\big]
\end{equation} 
for some $\varepsilon_0[C](y)$ are the following
\begin{equation}\label{cyclic}
\mathscr{F}_0(X,Y,Z)+\mathscr{F}_0(Y,Z,X)+\mathscr{F}_0(Z,X,Y)=0\,,
\end{equation}
\begin{equation}\label{factor}
\partial_X \partial_Y\Big(e^{-i(X+Y)}\mathscr{F}_0(X,Y,Z)\Big)=0\,,
\end{equation}
\begin{equation}\label{strange}
\big(\partial_X+\partial_Y+\partial_Z\big)\mathscr{F}_0(X,Y,Z)=i\mathscr{F}_0(X,Y,Z)\,.
\end{equation}
Here we have used notation from the section \ref{Explicit_f}
\begin{equation}
X:=y^\alpha p_{1\alpha}\,,\;\; Y:=y^\alpha p_{2\alpha}\,,\;\; Z:=p_{1\alpha} p_2 {}^\alpha\,.
\end{equation}
After star-product computation using the exponential form (cf. \eqref{exponential_form})  gauge transformation \eqref{gauge_0} can be brought to a form
\begin{equation}\label{basis}
\mathscr{F}_0(X,Y,Z)=e^{iY}\Phi(X-Z)-e^{iX}\Phi(Z-Y)\,.
\end{equation}
Performing cyclic permutation with $X,Y$ and $Z$ 
\begin{equation}
(X,Y,Z)\rightarrow (Y,Z,X)\rightarrow (Z,X,Y)
\end{equation}
from \eqref{basis} we get 3 equations that can be written as
\begin{equation}\label{system}
\left(\begin{matrix}
e^{iY} & 0 & -e^{iX} \\
-e^{iY} & e^{iZ} & 0\\
0 & -e^{iZ} & e^{iX}
\end{matrix}\right)\cdot \left(\begin{matrix}
\Phi(X-Z)\\
\Phi(Y-X)\\
\Phi(Z-Y)
\end{matrix}\right)=\left(\begin{matrix}
\mathscr{F}_0(X,Y,Z)\\
\mathscr{F}_0 (Y,Z,X)\\
\mathscr{F}_0 (Z,X,Y)
\end{matrix}\right)\,.
\end{equation}
Summing up all three equations from the system we obtain cyclic constraint on $\mathscr{F}_0(X,Y,Z)$ \eqref{cyclic}.

Once \eqref{cyclic} is fulfilled one can formally solve \eqref{system} as a linear system of equations. Namely, we treat \eqref{system} as 
\begin{equation}\label{system2}
\left(\begin{matrix}
e^{iY} & 0 & -e^{iX} \\
-e^{iY} & e^{iZ} & 0\\
0 & -e^{iZ} & e^{iX}
\end{matrix}\right)\cdot \left(\begin{matrix}
\phi_1(X,Y,Z)\\
\phi_2(X,Y,Z)\\
\phi_3(X,Y,Z)
\end{matrix}\right)=\left(\begin{matrix}
\mathscr{F}_0(X,Y,Z)\\
\mathscr{F}_0 (Y,Z,X)\\
\mathscr{F}_0 (Z,X,Y)
\end{matrix}\right)\,.
\end{equation}
and for fixed values of $X,Y$ and $Z$ obtain general solution in the form
\begin{equation}\label{phi_1}
\phi_1(X,Y,Z)=\frac{e^{-iY}}{3}\big(\mathscr{F}_0(X,Y,Z)-\mathscr{F}_0(Y,Z,X)\big)+\frac{e^{-iY}}{3}t(X,Y,Z)\,,
\end{equation}
\begin{equation}
\phi_2(X,Y,Z)=\frac{e^{-iZ}}{3}\big(\mathscr{F}_0(Y,Z,X)-\mathscr{F}_0(Z,X,Y)\big)+\frac{e^{-iZ}}{3}t(X,Y,Z)\,,
\end{equation}
\begin{equation}
\phi_3(X,Y,Z)=\frac{e^{-iX}}{3}\big(\mathscr{F}_0(Z,X,Y)-\mathscr{F}_0(X,Y,Z)\big)+\frac{e^{-iX}}{3}t(X,Y,Z)\,.
\end{equation}
Here part proportional to $t(X,Y,Z)$ represents the freedom in the solution of the homogeneous equation. We need to adjust $t(X,Y,Z)$ in the way that, for example, $\phi_1(X,Y,Z)$ does not depend on $Y$ and depends only on difference $(X-Z)$. Thus we have
\begin{equation}
\begin{cases}
\partial_Y\big[e^{-iY}\big(\mathscr{F}_0(X,Y,Z)-\mathscr{F}_0(Y,Z,X)\big)\big]+\partial_Y \big(e^{-iY}t(X,Y,Z)\big)=0\,,\\
\partial_Z\big[e^{-iZ}\big(\mathscr{F}_0(Y,Z,X)-\mathscr{F}_0(Z,X,Y)\big)\big]+\partial_Z\big(e^{-iZ}t(X,Y,Z)\big)=0\,,\\
\partial_X\big[e^{-iX}\big(\mathscr{F}_0(Z,X,Y)-\mathscr{F}_0(X,Y,Z)\big)\big]+
\partial_X\big(e^{-iX}t(X,Y,Z)\big)=0\,.
\end{cases}
\end{equation} 
Consistency of these PDE on $t(X,Y,Z)$, namely
\begin{equation}
(\partial_X\partial_Y-\partial_Y\partial_X)t(X,Y,Z)=(\partial_X\partial_Z-\partial_Z \partial_X)t(X,Y,Z)=(\partial_Y \partial_Z-\partial_Z\partial_Y)t(X,Y,Z)=0\,.
\end{equation}
brings factorization constraint \eqref{factor} on $\mathscr{F}_0(X,Y,Z)$. The generic solution for \eqref{factor} is of the form
\begin{equation}\label{factor_solution}
\mathscr{F}_0(X,Y,Z)=e^{iY}\Psi_1(X,Z)+e^{iX}\Psi_2(Y,Z)\,.
\end{equation}
The fact that $\phi_1(X,Y,Z)$ depends only on difference $(X-Z)$ brings another PDE on $t(X,Y,Z)$
\begin{equation}
\begin{cases}
(\partial_X+\partial_Z)\big(\mathscr{F}_0(X,Y,Z)-\mathscr{F}_0(Y,Z,X)\big)+(\partial_X+\partial_Z)t(X,Y,Z)=0\,,\\
(\partial_X+\partial_Y)\big(\mathscr{F}_0(Y,Z,X)-\mathscr{F}_0(Z,X,Y)\big)+(\partial_X+\partial_Y)t(X,Y,Z)=0\,,\\
(\partial_Y+\partial_Z)\big(\mathscr{F}_0(Z,X,Y)-\mathscr{F}_0(X,Y,Z)\big)+
(\partial_Y+\partial_Z)t(X,Y,Z)=0\,.
\end{cases}
\end{equation}
Using \eqref{system} and \eqref{system2} one finds $t(X,Y,Z)$ explicitly
\begin{equation}\label{t_solution}
2i t(X,Y,Z)=(\partial_X-\partial_Y)\mathscr{F}_0(X,Y,Z)+(\partial_Y-\partial_Z)\mathscr{F}_0(Y,Z,X)+(\partial_Z-\partial_X)\mathscr{F}_0(Z,X,Y)\,.
\end{equation}
After being plugged back, for example, in the first equation of \eqref{system2} one obtains additional constraint \eqref{strange} on $\mathscr{F}_0(X,Y,Z)$.

Eventually using $t(X,Y,Z)$ of the form \eqref{t_solution} from \eqref{phi_1} we can write down $\phi_1(X,Y,Z)$ as 
\begin{equation}\label{final_answer}
\phi_1(X,Y,Z)=\frac{i}{2} e^{-iY}\Big(e^{iX}\partial_X\big(e^{-iX}\mathscr{F}(X,Y,Z)\big)-e^{iZ}\partial_Z\big(e^{-iZ} \mathscr{F}(Y,Z,X)\big)\Big)\,.
\end{equation}
Expecting if \eqref{final_answer} depends only $(X-Z)$ one finds that this is indeed the case provided \eqref{cyclic},\eqref{factor},\eqref{strange} are fulfilled. Let us look closely at condition
\begin{equation}
(\partial_X+\partial_Z)\phi_1(X,Y,Z)=0\,.
\end{equation}
After plugging $\mathscr{F}_0(X,Y,Z)$ in the form \eqref{factor_solution} one obtains after simple algebra
\begin{equation}\label{Psis}
(\partial_X+\partial_Z)\Big[e^{iX}\partial_X\big(e^{-iX}\Psi_1(X,Z)\big)-e^{iZ}\partial_Z\big(e^{-iZ}\Psi_2(Z,X)\big)\Big]=0\,.
\end{equation}
Plugging \eqref{factor_solution} into \eqref{strange} it casts into  corresponding constraints on $\Psi_{1,2}$
\begin{equation}
e^{iY}(\partial_X+\partial_Z)\Psi_1(X,Z)+e^{iX}(\partial_Y+\partial_Z)\Psi_2(Y,Z)=0\,.
\end{equation}
Further massaging the expression one gets
\begin{equation}
-e^{-iY}(\partial_Y+\partial_Z)\Psi_2(Y,Z)=e^{-iX}(\partial_X+\partial_Z)\Psi_1(X,Z)
\end{equation}
Taking the derivative w.r.t. $X$ it casts into
\begin{equation}
\partial_X\big(e^{-iX}(\partial_X+\partial_Z)\Psi_1(X,Z)\big)=0\,
\end{equation}
which is equivalent to vanishing of $\Psi_1$-term in \eqref{Psis}. $\Psi_2$-term is treated analogously.

The only thing left to check is whether 
\begin{equation}
\phi_2(X,Y,Z)=\phi_1(Y,Z,X)\,,\, \phi_2(Y,Z,X)=\phi_3(X,Y,Z)\,,\ldots
\end{equation}
This is also true. Hence conditions \eqref{cyclic},\eqref{factor} and \eqref{strange} are necessary and sufficient for the change of the field frame to be a gauge transformation.


\end{document}